\documentclass[paper,notoc,nohyper]{JHEP3}
\usepackage{epsfig,cite}
\usepackage{amsbsy} 
\usepackage{epsfig,rotating}

\def\be{\begin{equation}}   
\def\ee{\end{equation}}
\def\bea{\begin{eqnarray}}  
\def\eea{\end{eqnarray}}

\def\O{y}
\def\asb{\left(\frac{\alpha_s}{2\pi}\right)}

\def\CA{C_A}

\def\mom#1{\langle #1 \rangle}

\def\d{\hbox{d}}

\title{\boldmath 
NNLO moments of event shapes in $e^+e^-$ annihilation
}
\author{
A.~Gehrmann--De Ridder\\
Institute for Theoretical Physics, ETH, CH-8093 Z\"urich,
Switzerland\\ 
E-mail: \email{gehra@phys.ethz.ch}}
\author{
T.~Gehrmann\\
Institut f\"ur Theoretische Physik, Universit\"at Z\"urich,
Winterthurerstrasse 190,\\ CH-8057 Z\"urich, Switzerland\\
E-mail: \email{thomas.gehrmann@physik.unizh.ch}}
\author{E.W.N.~Glover\\
Institute for Particle Physics Phenomenology, 
        Department of Physics,\\
        University of Durham, Durham, DH1 3LE, UK\\
	E-mail: \email{e.w.n.glover@durham.ac.uk}}
\author{G.~Heinrich\\
Institute for Particle Physics Phenomenology, 
        Department of Physics,\\
        University of Durham, Durham, DH1 3LE, UK\\
	E-mail: \email{gudrun.heinrich@durham.ac.uk}}

\abstract{
We compute the next-to-next-to-leading order (NNLO) QCD corrections to the first
five moments of six event shape variables related to three-particle final
states  in electron-positron annihilation; the  thrust, the heavy jet mass,  the
$C$-parameter, the wide and total jet broadenings and the  three-to-two-jet
transition parameter in the Durham algorithm  $Y_3$.  The NNLO corrections to
the first moment are moderate for all event shapes, while the  renormalisation
scale dependence of the theoretical prediction is substantially reduced compared
to the previously existing NLO results.  From a comparison with data from JADE
and OPAL, we observe that the energy dependence of the moments of the wide jet broadening and
$Y_3$ can be largely explained without any non-perturbative power corrections,
while the other observables exhibit a clear need for power-like contributions
at low centre-of-mass energy. 
}
\keywords{QCD, Jets, LEP and ILC Physics, NLO and NNLO Computations}
\preprint{{ZU-TH 04/09}, {IPPP/09/15}}

\begin{document}

\section{Introduction}
\label{sec:intro}

Event shape variables in $e^+e^-$ annihilation provide an ideal testing 
ground to study Quantum Chromodynamics (QCD) and have been 
measured and studied extensively in the last two decades. 
In particular, event shape variables are  interesting
for studying the interplay between perturbative and non-perturbative dynamics. 
Apart from distributions of these observables, to which the NNLO corrections 
recently have become available~\cite{GehrmannDeRidder:2007hr,GehrmannDeRidder:2007bj,GehrmannDeRidder:2007jk,Weinzierl:2009ms,Weinzierl:2009nz,weinzierlnew},
one can also study mean values and higher moments. 
The $n$th moment of an event shape observable $y$ is
defined by 
\begin{equation}
\mom{y^n}=\frac{1}{\sigma_{\rm{had}}}\,\int_0^{y_{\rm{max}}} y^n 
 \frac{\d\sigma}{\d y} \d y \;,
\end{equation}
where $y_{\mathrm{max}}$ is the kinematically allowed upper limit of the
observable.  
Moments were measured for a variety of different event shape variables in the 
past. The most common observables $y$ of three-jet type are: 
thrust $T$~\cite{farhi} (where 
moments of $y=(1-T)$ are taken), the heavy jet mass $\rho = M_H^2/s$~\cite{mh}, 
the $C$-parameter~\cite{c}, the wide and total jet broadenings $B_W$ and 
$B_T$~\cite{bwbt}, and the 
three-to-two-jet transition parameter in the Durham algorithm 
$Y_3$~\cite{durham}. 
Definitions for all observables are given in, for example, 
Ref.~\cite{GehrmannDeRidder:2007hr}.
Moments with $n\geq 1$ have been measured by several experiments, most extensively by 
{\small JADE}~\cite{MovillaFernandez:1997fr} and {\small OPAL}~\cite{Abbiendi:2004qz},
but also by {\small DELPHI}~\cite{Abreu:1999rc} and L3~\cite{Achard:2004sv}.
A combined analysis of JADE and OPAL results has been performed in Ref.~\cite{Kluth:2000km}.

As the calculation of moments involves an integration over the full phase
space,  they offer a way of comparing to data which is complementary to the use
of distributions, where in general cuts on certain kinematic regions are
applied. Furthermore, the two extreme kinematic limits -- two-jet-like events 
and multi-jet-like events -- enter with different weights in each moment:  the
higher the order $n$ of the moment, the more it becomes sensitive to the 
multi-jet region. Therefore it is particularly interesting to study the NNLO 
corrections to higher moments of event shapes, as these corrections  should
offer a better description of the multi-jet region due to the inclusion of 
additional radiation at parton level.

Moments of event shape variables can be divided into a perturbative 
and a non-perturbative part, 
\be
\mom{y^n}=\mom{y^n}_{\rm{pt}}+\mom{y^n}_{\rm{np}}\;, 
\label{mom}
\ee
where the non-perturbative part accounts for hadronisation effects.  The
non-perturbative part is suppressed by powers of $\lambda_p/Q^{p}\; (p\geq 1)$,
where  $Q\equiv \sqrt{s}$ is the centre of mass energy  and $\lambda_1$ is of
the order of  $\Lambda_{QCD}$.  The functional form of $\lambda_p$ has been
discussed quite extensively in the literature,  but as this parameter is closely
linked to  non-perturbative effects, it cannot be fully derived from first
principles. 

The power corrections can be related to infrared renormalons 
in the perturbative QCD expansion for the event shape 
variable~\cite{Manohar:1994kq,Webber:1994cp,Korchemsky:1994is,Dokshitzer:1995zt,Akhoury:1995sp,Dokshitzer:1995qm,Nason:1995hd,Dokshitzer:1997ew}. 
The analysis of infrared renormalon ambiguities suggests power corrections of the 
form $\lambda_p/Q^{p}$, but cannot make unique predictions for $\lambda_p$: 
it is only the sum of perturbative and non-perturbative contributions in 
(\ref{mom}) that becomes well-defined~\cite{Beneke:1998ui}.
Different ways to regularize the IR renormalon singularities 
have been worked out in the literature. 
One approach is to introduce an  IR cutoff $\mu_{I}$ and to replace the strong 
coupling constant below the scale $\mu_{I}$ by an effective coupling 
such that the integral of the coupling below $\mu_{I}$ has a finite value\cite{Dokshitzer:1995qm,Dokshitzer:1997ew}
\be
\frac{1}{\mu_I}\int_0^{\mu_I} dQ \,\alpha_{\rm{eff}}(Q^2)=\alpha_0(\mu_I)\;.
\label{alpha0}
\ee
This dispersive model for the strong coupling leads to a shift in the distributions
\be
\frac{\d\sigma}{\d y}=\frac{\d\sigma_{\rm{pt}}}{\d y}\,(y-a_y\,{\cal P})\;,
\ee
where the numerical factor $a_y$ depends on the event shape, while 
${\cal P}$ is believed to be universal  and scales with the CMS energy like  $\mu_I/Q$.
We refer the interested reader to Refs.~\cite{Dokshitzer:1997ew,Dokshitzer:1997iz,Dokshitzer:1998pt} for further details. 
 
Another ansatz is that of  Korchemsky et
al.~\cite{Korchemsky:2000kp,Belitsky:2001ij}  who suggest a shape function valid
equally for thrust, $C$-parameter  and heavy jet mass $\rho$,   which is
independent of the energy scale and  takes into account the energy flow into two
hemispheres of the final state. In the  region $y \gg \Lambda_{QCD}/Q$, their
predictions for the  mean values of $1-T$ and $\rho$ coincide with those of 
renormalon based models, i.e.\ the  leading power corrections are parametrised by
a single non-perturbative scale $\lambda_1$.  In the case of the $C$-parameter
however, they find an additional  contribution modifying the size of the $1/Q$
correction.   This formalism also allows one to study the region $y\sim
\Lambda_{QCD}/Q$, where power corrections of the form $(\Lambda_{QCD}/(Qy))^p$
for arbitrary $p$ become important. 

Gardi et al. \cite{Gardi:1999dq,Gardi:2000yh,Gardi:2001ny,Gardi:2002bg} make an
analysis  using the formalism of the so-called ``dressed gluon
exponentiation"\,\cite{Gardi:2001ny},   which resums the renormalon contribution
to the Sudakov exponent, taking into account  factorially enhanced subleading
logarithms. Within this formalism, power law corrections beyond the leading term
are predicted which probe the region  $y\sim \Lambda_{QCD}/Q$. In particular,
this formalism predicts that power corrections of the form  $\lambda_2/Q^2$ are
suppressed, which is also indicated by the data. Ref.\,\cite{Gardi:2002bg}
contains a detailed analysis of the power corrections for  the thrust and heavy
jet mass distributions,  taking also hadron mass effects into account and
deducing from renormalon analysis that the non-perturbative corrections to both 
distributions can be described by a single shape function.  In
Ref.~\cite{Gardi:2000yh},  characteristic functions describing the leading and
subleading power corrections to the first four moments of the thrust variable
are derived. 

In Ref.~\cite{Campbell:1998qw}, a renormalisation group improved (RGI) treatment
of the first moment of event shape observables is suggested and compared to data. 
A simultaneous fit of three parameters -- $\Lambda_{QCD}$, a parameter 
$\rho_2$ describing uncalculated perturbative 
higher order corrections and a parameter $K_0$ controlling the power corrections -- 
showed for the cases of $\langle 1-T\rangle$ and $\langle \rho \rangle$ that 
the data are consistent with $K_0=0$ within the RGI approach without leading to 
unreasonably large values of $\rho_2$, but  do not allow definite conclusions.

Overviews on the various approaches to estimate non-perturbative corrections 
can be found in Refs.~\cite{Beneke:1998ui,Dasgupta:2003iq}. 
Recently, significant progress in the theoretical description of power corrections 
has also been made using Soft-Collinear Effetive Theory (SCET\,\cite{Bauer:2001ct,Bauer:2001yt,Bauer:2002nz}), 
see e.g. Ref.~\cite{Bauer:2008dt}.

An extensive comparison of $e^+e^-$ data on events shapes   with NLO+NLLA QCD
predictions, supplemented by power corrections within the  approach of
\cite{Dokshitzer:1995qm,Dokshitzer:1997ew}, has been performed in 
\cite{MovillaFernandez:2001ed}. A determination of $\alpha_s$ and $\alpha_0$
from fits to distributions and mean values  of five event shapes has been
carried out, and indications were found that uncalculated  higher orders may
contribute significantly to the non-perturbative corrections. A determination of
$\alpha_s$ based on moments of event shapes  calculated up to NLO has recently
been reported in  \cite{Pahl:2008uc}. 

Based on the NNLO corrections published in~\cite{GehrmannDeRidder:2007hr}, 
several new determinations of  the strong coupling constant have become
available  recently. In  Ref.~\cite{Dissertori:2007xa}, fits of six event shape
distributions to  {\small ALEPH} data have been made.  In  \cite{Bethke:2008hf},
the same perturbative results have been fitted to {\small JADE}
data, including also matching~\cite{Gehrmann:2008kh}  onto the resummation of
large logarithms~\cite{resumall}  in the next-to-leading log approximation
(NLLA).

A very recent study of non-perturbative contributions to the thrust
distribution,  based on the NNLO corrections published
in~\cite{GehrmannDeRidder:2007hr}  and including  NLLA resummation, as well as
power corrections  within the low-scale effective coupling
formalism~\cite{Dokshitzer:1995qm,Dokshitzer:1997ew},  can be found
in~\cite{Davison:2008vx}.  This work also provides a determination of 
$\alpha_s$ and $\alpha_0$ based on fits to a wealth of experimental data for
the  thrust distribution and  confirms the conjecture  of
\cite{MovillaFernandez:2001ed} that the explicit $1/Q$ corrections become
smaller as higher perturbative orders are included. This corroborates the
hypothesis that the divergent renormalon contribution  can be regularized by
modifying the strong coupling at low scales, leading to the effective coupling 
used in eq.~(\ref{alpha0}).

In this paper, we provide the NNLO corrections to the first five moments of  the
event shapes thrust $T$, the normalised heavy jet mass $\rho$,  the
$C$-parameter,  the wide and total jet broadenings $B_W$ and $B_T$ and the
transition from three-jet to  two-jet final states in the Durham jet algorithm
$Y_3$. After  reviewing the theoretical framework and fixing the notation in 
Section~\ref{sec:th}, we present results for the NNLO coefficients of the  first
five moments of the above-mentioned event shape distributions in 
Section~\ref{sec:coeff}. The numerical impact of these corrections and
implications for the description of experimental event shape data are discussed
in Section~\ref{sec:num}.

\section{Theoretical framework}
\label{sec:th}
The perturbative expansion for the distribution of a 
generic event shape observable $\O$ up to NNLO at centre-of-mass energy 
$\sqrt{s}$ 
for renormalisation scale $\mu^2 = s$ and 
$\alpha_s\equiv \alpha_s(s)$  is given by
\begin{eqnarray}
\frac{1}{\sigma_{{\rm had}}}\, \frac{\d\sigma}{\d \O} &=& 
\left(\frac{\alpha_s}{2\pi}\right) \frac{\d \bar A}{\d \O} +
\left(\frac{\alpha_s}{2\pi}\right)^2 \frac{\d \bar B}{\d \O}
+ \left(\frac{\alpha_s}{2\pi}\right)^3 
\frac{\d \bar C}{\d \O} + {\cal O}(\alpha_s^4)\;.
\label{eq:NNLO}
\end{eqnarray}
Here the event shape distribution  
is normalised to the total hadronic cross section $\sigma_{\rm{had}}$.
With the assumption of massless quarks, then   
at NNLO 
we have,
  \begin{equation}
  \sigma_{\rm{had}}=\sigma_0\,
\left(1+\frac{3}{2}C_F\,\left(\frac{\alpha_s}{2\pi}\right)
+K_2\,\left(\frac{\alpha_s}{2\pi}\right)^2+{\cal O}(\alpha_s^3)\,
\right) \;,
\end{equation}
where the Born cross section for $e^+e^- \to q \bar q$ is
\begin{equation}
\sigma_0 = \frac{4 \pi \alpha}{3 s} N \, e_q^2\;.
\end{equation}
The constant $K_2$ is given by~\cite{Chetyrkin:1979bj,Celmaster:1979xr,Dine:1979qh}
\begin{equation}
  K_2=\frac{1}{4}\left[- \frac{3}{2}C_F^2
+C_FC_A\,\left(\frac{123}{2}-44\zeta_3\right)+C_FT_RN_F\,(-22+16\zeta_3)
 \right] \;,
\end{equation}
where the QCD colour factors are
\begin{equation}
\CA = N,\qquad C_F = \frac{N^2-1}{2N},
\qquad T_R = \frac{1}{2}\; 
\end{equation}
for $N=3$ colours and $N_F$ light quark flavours.

The  perturbative QCD expansion of $\mom{y^n}$ is then given by
\begin{equation}
  \mom{y^n}(s,\mu^2 = s) = \asb \bar{{\cal A}}_{y,n}  + 
\asb^2 \bar{{\cal B}}_{y,n} + \asb^3 \bar{{\cal C}}_{y,n} +
  {\cal O}(\alpha_s)^4\;.
\label{eq:pertexp}
\end{equation}

In practice, we compute the perturbative coefficients ${\cal A}_n$, 
${\cal B}_n$ and ${\cal C}_n$, which are 
all normalised to 
$\sigma_0$:
\begin{eqnarray}
\frac{1}{\sigma_0}\,\int_0^{y_{{\rm max}}} \d \O \, \O^n \,
 \frac{\d\sigma}{d \O} &=& 
\left(\frac{\alpha_s}{2\pi}\right) {\cal A}_{y,n} +
\left(\frac{\alpha_s}{2\pi}\right)^2 {\cal B}_{y,n}
+ \left(\frac{\alpha_s}{2\pi}\right)^3 
{\cal C}_{y,n}  + {\cal O}(\alpha_s^4)\,.
\label{eq:NNLOsigma0}
\end{eqnarray}
${\cal A}_{y,n}$, 
${\cal B}_{y,n}$ and ${\cal C}_{y,n}$ 
are straightforwardly related to $\bar{{\cal A}}_{y,n}$, 
$\bar{{\cal B}}_{y,n}$ 
and $\bar{{\cal C}}_{y,n}$ by
\begin{eqnarray}
\bar{{\cal A}}_{y,n} &=& {\cal A}_{y,n}\;,\nonumber \\
\bar{{\cal B}}_{y,n} &=& {\cal B}_{y,n} 
- \frac{3}{2}C_F\, {\cal A}_{y,n}\;,\nonumber \\
\bar{{\cal C}}_{y,n} &=& {\cal C}_{y,n} 
-  \frac{3}{2}C_F\,{\cal B}_{y,n}
+ \left(\frac{9}{4}C_F^2\,-K_2\right)\,{\cal A}_{y,n} 
\;.\label{eq:ceff}
\end{eqnarray} 
These coefficients are computed at a renormalisation scale fixed to 
the centre-of-mass energy, 
and are therefore just dimensionless numbers for each observable and 
each value of $n$.

The computation of the coefficients is carried out using 
the  parton-level event generator program 
{\tt EERAD3}~\cite{GehrmannDeRidder:2007jk,GehrmannDeRidder:2008ug}, which
contains the relevant matrix elements with up to five external
partons~\cite{3jme,muw2,V4p,tree5p}, combined using an infrared
antenna subtraction method~\cite{ourant}. A recently discovered
inconsistency in the treatment of large-angle soft
radiation terms~\cite{weinzierlnew} in the original {\tt EERAD3}
implementation has been corrected. 
 They account for an initially observed
discrepancy between the {\tt EERAD3} results and the logarithmic
contributions (computed within SCET) to the thrust distribution to
 NNLO~\cite{Becher:2008cf},
which are now in full agreement.

In terms of the running coupling $\alpha_s(\mu^2)$, the 
NNLO expression for an event shape moment measured at centre-of-mass 
energy squared $s$ becomes:
\begin{eqnarray}
\mom{y^n} (s,\mu^2) &=& 
\left(\frac{\alpha_s(\mu)}{2\pi}\right) \bar{{\cal A}}_{y,n} +
\left(\frac{\alpha_s(\mu)}{2\pi}\right)^2 \left( 
\bar{{\cal B}}_{y,n}+ \bar{{\cal A}}_{y,n} \beta_0 
\log\frac{\mu^2}{s} \right)
\nonumber \\ &&
+ \left(\frac{\alpha_s(\mu)}{2\pi}\right)^3 
\bigg(\bar{{\cal C}}_{y,n}+ 2 \bar{{\cal B}}_{y,n}
 \beta_0\log\frac{\mu^2}{s}
+ \bar{{\cal A}}_{y,n} \left( \beta_0^2\,\log^2\frac{\mu^2}{s}
+ \beta_1\, \log\frac{\mu^2}{s}   \right)\bigg)  
\nonumber \\ &&
 + {\cal O}(\alpha_s^4)\;.
\label{eq:NNLOmu} 
\end{eqnarray}

\section{NNLO contributions}
\label{sec:coeff}
The perturbative LO and NLO coefficients ${\cal A}_{y,n}$, ${\cal B}_{y,n}$
have been known for many years~\cite{ert1,ert2,kunszt,cs}, and 
have been used extensively in the experimental studies of LEP data. 
Our major new results presented here are the perturbative 
NNLO coefficients ${\cal C}_{y,n}$. These coefficients 
receive contributions from six different colour factors:
$$ N^2\,,\quad N^0\,,\quad 1/N^2\,,\quad N_F\,N\,,\quad N_F/N\,,\quad N_F^2\,.
$$
We have computed these six different contributions separately. 

The evaluation of the perturbative coefficients  ${\cal A}_{y,n}$, ${\cal
B}_{y,n}$ and ${\cal C}_{y,n}$ requires the  numerical integration of the
weighted perturbative event shape distributions  down to the exact two-jet
boundary $y=0$. At this boundary, all  event shape distributions diverge like
$1/y$, and are further enhanced  by large logarithmic corrections at NLO and
NNLO. Since the integrand for the $n$-th moment is  the distribution weighted by
$y^n$, the moments themselves are finite. The first moment does however contain
an integrable logarithmic  singularity, and thus receives sizable contributions
from the region $y\to 0$.

The numerical convergence of the perturbative coefficients computed  by {\tt
EERAD3} deteriorates for $y\to 0$ for the following reason.  The  four-particle
and five-particle phase spaces in {\tt EERAD3} are  generated with a lower
cut-off $\delta_0$ on all  dimensionless invariants $s_{ij}/s$. This cut-off  is
required to avoid sampling kinematical regions where  the numerical evaluation
of the four-parton one-loop matrix element  becomes unstable due to the presence
of large inverse Gram determinants,  and also to avoid regions where the antenna
subtraction procedure induces  cancellations over too many orders of magnitude
for the remainder to  be evaluated reliably. The independence of the results on 
$\delta_0$ serves  as a check for the proper implementation of the subtraction.
In our calculation of the event shape distributions~\cite{GehrmannDeRidder:2007hr}, we used $\delta_0=10^{-5}$ as a
default, observing that this value allowed  a reliable evaluation of all
contributions while ensuring numerical stability  and convergence. In
approaching $y\to \delta_0$ from above, the unresolved invariants  are no longer
small compared to the resolved invariants, and the calculation is likely to miss
parts of the finite remainder from the subtraction. For $y<\delta_0$, no
reliable evaluation is possible. Consequently, we have to impose a lower cut
$y_0$ on $y$ when  evaluating the perturbative coefficients. This cut is likely
to affect  especially the first moment. Due to the strong suppression of the 
$y\to 0$ region in the higher moments, this cut does not play a role  for $n\geq
2$. 

Therefore, in evaluating the perturbative coefficients,  especially for the
first moment, we carried out systematic studies to determine the values for the
technical parameters $\delta_0$ and $y_0$. It is observed that $y_0=10^{-5}$ is
sufficient for all observables.  In approaching this value, one observes already
a  reasonable convergence of all colour factor contributions to the first
moments, and from studies with lower values of $y_0 < 10^{-5}$, we estimate  the
contribution from this region to each colour factor  not  to exceed five per
cent of the total. The choice of $\delta_0$ depends on the event shape variable
under consideration. Already at NLO, one observes  that $\delta_0=y_0/10$ is
sufficiently small for a reliable evaluation  of the first moment of $(1-T)$,
$\rho$, $C$ and $Y_3$, while $B_W$ and  $B_T$ require much smaller $\delta_0$.
In the distribution of both  broadenings, one observes that a cut $\delta_0$
induces a modification  of the distributions up to $y\approx \sqrt{\delta_0}$,
thereby  rendering the numerical evaluation of the first moment of the 
broadenings much more challenging. This effect is illustrated in 
Figure~\ref{fig:bwnlo} for the NLO-contribution to the $B_W$ distribution,
evaluated for different values of $\delta_0$. 
For the NNLO contributions to  the first
moment of the broadenings, we chose $\delta_0 =10^{-7}$, which is  clearly a
compromise between the accuracy of the evaluation and the numerical 
convergence. We have verified for both broadenings  that for fixed
$y_0=10^{-5}$, the numerical values of  first moments do stabilize at $\delta_0
=10^{-7}$, and we estimate that the  individual colour factor contributions do
not change by more than three per cent for lower $\delta_0$. Full studies with
much lower $\delta_0$ were however not possible because of  rapid deterioration
of the numerical convergence.  For these reasons, a systematic error of about
50\% on the first moments  of the broadenings should be taken into account. 

In Tables~\ref{tab:col1} and~\ref{tab:col2}, we list the individual 
colour factor contributions to the first five moments of the different 
event shape variables. Table~\ref{tab:mom} contains the 
total LO, NLO and NNLO  coefficients. 
The moments given in Tables  \ref{tab:col1}
to \ref{tab:mom}
have been calculated with 
$\delta_0=10^{-7}$ for $B_W$ and $B_T$,
where studies of other $\delta_0$ values lead us to 
a systematic error estimate of 10\% (except for the first moment, where 
the systematic error is estimated at 50\%). 
For the other event shape variables, the default for the first moments 
is $\delta_0=10^{-6}$ with a systematic error of 6\%, while the values 
for the higher moments have been calculated with $\delta_0=10^{-5}$
and studies of other $\delta_0$ values result in
a systematic error estimate of 3\% for $(1-T)$, $\rho$, $C$ and $Y_3$.

The behaviour in the different event shape variables is best seen by
considering the ratios 
\begin{equation}
K_{\rm{NLO}}=\frac{\bar{{\cal A}}_{y,n}+\bar{{\cal B}}_{y,n}\,\left(\frac{\alpha_s}{2\pi}\right)}{\bar{{\cal A}}_{y,n}}
\end{equation}
 and
\begin{equation}
K_{\rm{NNLO}}=\frac{\bar{{\cal A}}_{y,n}+\bar{{\cal B}}_{y,n}\,\left(\frac{\alpha_s}{2\pi}\right)+{\bar{\cal C}}_{y,n}\,\left(\frac{\alpha_s}{2\pi}\right)^2}{\bar{\cal A}_{y,n}}\;, 
\end{equation}
which we display for $\alpha_s=0.124$ in Figures
\ref{fig:TMH}--\ref{fig:BWY3}. Although substantially above the 
world average, it turned out that this value of $\alpha_s$ is the best-fit
result~\cite{Dissertori:2007xa}
 of fitting the pure NNLO predictions to event shape data 
from ALEPH. Consequently, we also employ this value in the event shape 
studies here.
We observe that the perturbative corrections 
affect the different shape variables in a substantially different manner, 
which was already evident at NLO (see for example~\cite{Abbiendi:2004qz}).
While the size of the corrections is largely independent of $n$ for 
$\rho$, $B_W$ and $Y_3$ (and roughly ${\cal K}\approx 1.3$ 
at NNLO for all three variables), 
it increases with $n$ for $(1-T)$, $C$ and $B_T$ (amounting to 
${\cal K}\approx 3$ for $n=5$ for all three variables).
This picture is consistent with the observations made on the 
perturbative stability of the distributions~\cite{GehrmannDeRidder:2007hr}, 
where the former three displayed uniform and moderate 
NNLO corrections, resulting 
in a stabilization of the perturbative expansion, while the latter three suffered 
large corrections.

\clearpage

\section{Numerical results and comparison to data}
\label{sec:num}

Moments of event shapes have been measured by various $e^+e^-$ collider 
experiments at centre-of-mass energies ranging from 10~GeV to 206~GeV 
\cite{Pahl:2007zz}. These   data are  particularly interesting in view of the
determination of non-perturbative  power corrections. Since the power correction
contributions  scale like inverse powers of the centre-of-mass energy $Q$,  one
needs a large range in  energies to empirically disentangle their contribution
from the  perturbative contribution, which scales logarithmically with the
energy. 

However, even based on data covering about a factor twenty in energy range, 
missing higher order contributions may be misidentified as power corrections. 
The accuracy on the extraction of power correction contributions from data  is
therefore inherently linked to the precision of the perturbative prediction. We
therefore expect the NNLO corrections computed here to  lead to a considerable
improvement of the power correction  extraction  from event shape moments. 

In Figure \ref{fig:tn1}, we computed the energy dependence of the first
moments of all event shape variables for $\alpha_S(M_Z)=0.124$ at LO, NLO and 
NNLO. The theory uncertainty at each order is estimated by varying the 
renormalization scale in the strong coupling constant in the range
$Q/2<\mu_R<2\,Q$. The theory predictions are compared to data from 
JADE and OPAL \cite{Pahl:2007zz}.  
It can be clearly seen that the inclusion of the NNLO corrections
leads to a substantial reduction (by a factor two to three
compared to NLO) 
of the theoretical uncertainty in all 
variables. The theory error bands at NLO and NNLO generally overlap 
(although the overlap is only marginal for $(1-T)$, $C$ and $B_T$,
which have the largest corrections, see above). 
We observe that the energy dependence of the first moment is well 
described by the purely perturbative NNLO QCD contribution for $Y_3$, while
all other observables require the presence of 
additional non-perturbative contributions at low centre-of-mass
energies. Comparing NLO and NNLO predictions, we observe that 
the NNLO contribution comes closer to the experimental data at low 
$Q$ by about 30\% for $1-T$ and $C$, while only little improvement over NLO 
is observed for the other observables. The low-$Q$ discrepancy between data 
and NNLO theory is most pronounced in $\rho$, $C$ and $(1-T)$. 
It is especially interesting to note
that the required size of the power correction contribution is uncorrelated 
with size of the perturbative higher order corrections: both $\rho$ and 
$Y_3$ receive only moderate NLO and NNLO corrections; to describe the 
low-energy data on the first moment of $\rho$, one must postulate substantial
(larger than in any other observable) power corrections, while data on 
$Y_3$ are well described without power corrections. 

The higher moments of the six standard event shapes are displayed in  Figures
\ref{fig:tn2to5} ($1-T$), \ref{fig:rhon2to5} ($\rho$), \ref{fig:cn2to5} ($C$),
\ref{fig:y3n2to5} ($Y_3$), \ref{fig:btn2to5} ($B_T$) and \ref{fig:bwn2to5}
($B_W$). The predictions are compared to JADE and OPAL  data 
\cite{Pahl:2007zz}. The qualitative behaviour of the higher moments  of the
different shape variables is similar to what was observed for the first
moments. For $(1-T)$, $C$ and $B_T$, we again observe that the NLO  and NNLO
theory error bands merely overlap. Despite the increase in the  size of the NNLO
corrections to the higher moments, this situation does not deteriorate for
higher $n$,  mainly because of the equally large increase of the NLO corrections.
This behaviour should however be taken as an indication of the poor
perturbative convergence of these three observables. For all $n$, we observe a
substantial discrepancy between NNLO theory and experimental data at low-$Q$ 
in   $\rho$, $C$ and $(1-T)$ and $B_T$, indicating large power corrections to
these  observables. The relative size of the discrepancy does not decrease with
$n$,  illustrating the relevance of power corrections throughout the full 
kinematical range of the event shape variable (and not only in the two-jet
region). Data on the higher moments of $B_W$ and $Y_3$ are well  described by
NNLO theory, leaving only little room for power correction contributions to
these observables.  

With the new NNLO corrections, the theory uncertainty  is reduced to the few per
cent level  for the moments of all event shape observables. It should therefore
now be  possible to perform precise studies of power corrections on these data 
sets.

\section{Conclusions and Outlook}
\label{sec:conc}

In this paper, we have computed the NNLO corrections to moments of  the six most
commonly studied event shape variables: $(1-T)$, $\rho$, $C$,  $B_T$, $B_W$ and
$Y_3$. Especially the first moment  receives substantial contributions from
kinematical configurations deep  inside the two-jet region, well beyond the
region probed by event shape distributions. Consequently, the evaluation of the
first moment is technically  demanding, in particular for the jet broadenings. 

The NNLO corrections to the first moment are moderate for all event shapes,
leading to a  ${\cal K}$-factor of order unity (in the  range 0.95 to 1.4 for
the different shapes). For higher moments, we  observed that the NNLO
corrections to the moments of $(1-T)$, $C$ and  $B_T$ increase with increasing
moment number $n$, resulting in large NNLO  ${\cal K}$-factors of up to 3. For
$B_W$, $\rho$ and $Y_3$, we observe that  the corrections remain moderate
towards higher $n$, indicating the better perturbative stability of these three
variables. 

The energy dependence of the event shape moments can be used for the 
empirical determination of power corrections due to non-perturbative effects. 
For these studies, a reliable description of the perturbative contributions 
is mandatory. We observe that the newly computed NNLO corrections lead to
a substantial reduction of the theoretical uncertainty on these predictions. 
From a comparison with data from JADE and OPAL, we observe that the energy
dependence of the moments of $B_W$ and $Y_3$ can be largely described without 
any power corrections, while the other observables show clear indications 
for power-like contributions at low centre-of-mass energy. 

In view of the newly computed NNLO perturbative corrections, it will be  very
interesting to revisit the existing  non-perturbative power correction models.
Given the considerably reduced  uncertainty on the perturbative prediction, a
more precise extraction of hadronization parameters can now be envisaged.

\section*{Acknowledgements}

AG and TG would like to acknowledge the Aspen Center of Physics,  where part of
this work was carried out.  GH would like to thank the ITP, University of
Z\"urich for hospitality. EWNG gratefully acknowledges the support of the
Wolfson Foundation and the Royal Society. This research was supported in part by
the Swiss National Science Foundation (SNF) under contracts PP0022-118864 and
200020-117602, by the UK Science and Technology Facilities Council and  by the
European Commission's Marie-Curie Research Training Network under contract
MRTN-CT-2006-035505 ``Tools and Precision Calculations for Physics Discoveries
at Colliders''.

\bibliographystyle{JHEP}

\begin{sidewaystable}[t]
{
\begin{tabular}{|lr|c|c|c|c|c|c|}
\hline \rule[-4mm]{0mm}{10mm}
& & $\langle (1-T)^n \rangle$ &  $\langle \rho^n \rangle$ & 
 $\langle C^n \rangle$ &  $\langle B_W^n \rangle$ & 
 $\langle B_T^n \rangle$ & $\langle Y_3^n \rangle$  \\[1mm] \hline
\rule[0mm]{0mm}{2.5ex}
$N^2$ & $n=1$ 
      & $2301 \pm 11$     & $993 \pm 12$    &$8816\pm 37$& $348 \pm 34$    & $3007 \pm 59$ & $470\pm 5$  \\
& $2$ & $337.1 \pm 0.6$  & $120.5\pm 0.6$  & $4413\pm 5$& $174 \pm1$      &$745\pm 1$ & $45.6\pm 0.3$ \\
& $3$ & $63.2 \pm 0.2$   & $15.7 \pm 0.1$  & $2381\pm 2$&$23.2 \pm 0.2$   &$145.6\pm 0.2$ & $6.64\pm  0.07$  \\
& $4$ & $13.94 \pm 0.02$ & $2.42\pm 0.03$  & $1471\pm1$ &$3.20\pm 0.03$   &$32.13\pm 0.03$ & $1.20 \pm 0.01$\\ 
& $5$ & $3.490 \pm 0.005$& $0.438\pm 0.006$& $ 984\pm2$ &$0.456\pm 0.005$ &$7.86 \pm 0.01$ & $0.249 \pm   0.002$ \\[1mm]
\hline 
\rule[0mm]{0mm}{2.5ex} $N^0$ & $n=1$ 
     & $34.68 \pm 0.12$  & $85.6 \pm 0.2$        & $285.4\pm 0.4$       &$4.0\pm 4.4 $& $383 \pm 2 $& $14.38\pm  0.06$  \\
&$2$ & $ -19.62 \pm 0.01$ & $-2.10 \pm 0.01$     & $-220.1\pm 0.2$ &$7.66\pm 0.02 $& $-21.64 \pm 0.03 $& $-1.181\pm  0.005 $  \\
&$3$ & $-4.630 \pm 0.002$ & $-0.564 \pm 0.003$   & $-165.13\pm 0.06$&$-0.028\pm 0.002 $ & $-8.656 \pm 0.005 $& $-0.243 \pm 0.002$\\
&$4$ &$-1.1240 \pm 0.0006$ & $-0.0973 \pm 0.0009$& $-113.92\pm 0.03$&$-0.0474\pm 0.0007$& $-2.285 \pm 0.002 $&$-0.0494 \pm 0.0005$ \\
&$5$ &$-0.2938 \pm 0.0001$ & $-0.0163 \pm 0.0003$& $ -80.72\pm 0.03$&$-0.2938\pm 0.0001$& $-0.6025 \pm 0.0005 $& $-0.0102\pm 0.0001$\\[1mm]
\hline
\rule[0mm]{0mm}{2.5ex} $1/N^2$ & $n=1$
     & $0.202 \pm 0.008 $ &  $7.51\pm 0.03$ & $3.31 \pm 0.03 $          & $85.6 \pm 0.2$ & $16.17 \pm 0.05$ & $0.5788 \pm      0.0041$\\
&$2$ & $ -0.2270 \pm 0.0008 $&$ -0.0799 \pm 0.0009$ & $-4.900 \pm 0.007$& $0.232 \pm   0.001$ & $-2.620\pm 0.004$ & $-0.0205 \pm     0.0007$\\
&$3$ & $ -0.0010 \pm 0.0002 $&$-0.0133 \pm 0.0003 $ & $-0.931 \pm 0.006$& $-0.0089\pm  0.0002$ & $-0.3140 \pm 0.0006$ & $-0.00432 \pm   0.0001$\\
&$4$ & $0.00458 \pm 0.00004 $&$ -0.00257\pm  0.00006$ & $0.044\pm 0.004$& $-0.00196\pm 0.00003$ & $-0.0467 \pm 0.0001$ & $-0.00077 \pm    0.00004$\\
&$5$ & $ 0.00197 \pm  0.00001$&$-0.00065\pm 0.00002 $& $0.308 \pm 0.003$& $-0.00036 \pm 0.00001$ & $-0.00790 \pm 0.00003$ & $-0.00019 \pm 0.00001$\\[1mm]
\hline
\end{tabular}
\caption{Individual colour factor contributions to event shape moments 
at NNLO: $N^2$, $N^0$, $1/N^2$.
The given errors are the statistical errors from the numerical integration. Systematic errors originating from technical cut parameters (see text) 
are not shown in the table. \label{tab:col1}}}
\end{sidewaystable} 

\begin{sidewaystable}[t]
{
\begin{tabular}{|lr|c|c|c|c|c|c|}
\hline \rule[-4mm]{0mm}{10mm}
& & $\langle (1-T)^n \rangle$ &  $\langle \rho^n \rangle$ & 
 $\langle C^n \rangle$ &  $\langle B_W^n \rangle$ & 
 $\langle B_T^n \rangle$ & $\langle Y_3^n \rangle$  \\[1mm] \hline
\rule[0mm]{0mm}{2.5ex}
$N_F\, N$ & $n=1$ 
      & $-1716\pm 18$     & $-1085 \pm15$    &$-6930 \pm80$& $-664 \pm33$    &$-3279 \pm 41$ & $-451 \pm 8$  \\
& $2$ & $-164 \pm 1$  & $-95.8 \pm 0.9$  & $-2244 \pm11$& $-172 \pm 2$      &$-434 \pm 2$ & $-36.7 \pm 0.5$ \\
& $3$ & $-26.1 \pm 0.2$   & $-12.8 \pm 0.2$  & $-1034 \pm 6$&$-21.6 \pm 0.2$   &$-70.6 \pm 0.3$ & $-5.42 \pm 0.08$  \\
& $4$ & $-5.23 \pm 0.04$ & $-2.30 \pm 0.04$  & $-572 \pm 4$ &$-3.21 \pm 0.04$   &$-14.06 \pm 0.06$ & $-1.05 \pm 0.02$\\ 
& $5$ & $-1.202 \pm 0.008$& $-0.47 \pm 0.01$& $-351 \pm 3$ &$-0.56 \pm 0.01$ &$-3.23 \pm 0.02$ & $-0.218 \pm 0.004$ \\[1mm]
\hline 
\rule[0mm]{0mm}{2.5ex}
$N_F/N$ & $n=1$ 
      & $-6.67 \pm 0.01$     & $-67.92 \pm 0.11$    &$ -85.48 \pm 0.07$& $-465.2 \pm 0.3$    & $-182.6 \pm 0.2$ & $-6.33 \pm 0.02$  \\
& $2$ & $7.74 \pm 0.02$  & $1.39 \pm 0.02$  & $102.9 \pm 0.1$& $-3.16 \pm 0.02$      &$23.79 \pm 0.03$ & $0.642 \pm 0.007$ \\
& $3$ & $1.478 \pm 0.004$   & $0.312 \pm 0.003$  & $59.29 \pm 0.09$&$0.133 \pm 0.003$   &$4.903 \pm 0.008$ & $0.123 \pm 0.001$  \\
& $4$ & $0.3179 \pm 0.0005$ & $0.0596 \pm 0.0006$  & $36.22 \pm 0.05$ &$0.0398 \pm 0.0004$   &$1.075 \pm 0.002$ & $0.0241 \pm 0.0004$\\ 
& $5$ & $0.0763 \pm 0.0003$& $0.0129 \pm 0.0002$& $23.49 \pm 0.03$ &$0.0072 \pm 0.0001$ &$0.2565 \pm 0.0005$ & $0.0052 \pm 0.0002$ \\[1mm]
\hline
\rule[0mm]{0mm}{2.5ex}
$N_F^2$ & $n=1$ 
      & $253.9 \pm 0.2$     & $271.32 \pm 0.19$    &$1123.5 \pm 0.5$& $738.3 \pm 0.3$    & $710.4 \pm 0.2$ & $83.24 \pm 0.05$  \\
& $2$ & $12.14 \pm 0.04$  & $15.00 \pm 0.04$  & $173.2 \pm 0.4$& $33.64 \pm 0.07$      &$27.79 \pm 0.05$ & $5.50 \pm 0.02$ \\
& $3$ & $1.388 \pm 0.007$   & $2.001 \pm 0.008$  & $55.8 \pm 0.2$&$3.829 \pm 0.007$   &$2.42 \pm 0.01$ & $0.792 \pm 0.003$  \\
& $4$ & $0.213 \pm 0.002$ & $0.364 \pm 0.001$  & $22.2 \pm 0.1$ &$0.609 \pm 0.002$   &$0.240 \pm 0.002$ & $0.1479 \pm 0.0008$\\ 
& $5$ & $0.0368 \pm 0.0004$& $0.0760 \pm 0.0004$& $9.15 \pm 0.07$ &$0.1129 \pm 0.0004$ &$0.0111 \pm 0.0005$ & $0.0320 \pm 0.0003$ \\[1mm]
\hline
\end{tabular}
\caption{Individual colour factor contributions to event shape moments 
at NNLO: $N_F\, N$, $N_F/N$, $N_F^2$.
The given errors are the statistical errors from the numerical integration. Systematic errors originating from technical cut parameters (see text) 
are not shown in the table.\label{tab:col2}\hspace{4cm}}}
\end{sidewaystable}

\begin{sidewaystable}[t]
\begin{tabular}{|lr|c|c|c|c|c|c|}
\hline \rule[-4mm]{0mm}{10mm}
& & $\langle (1-T)^n \rangle$ &  $\langle \rho^n \rangle$ & 
 $\langle C^n \rangle$ &  $\langle B_W^n \rangle$ & 
 $\langle B_T^n \rangle$ & $\langle Y_3^n \rangle$  \\[1mm] \hline
\rule[0mm]{0mm}{2.5ex}
${\cal A}$ & $n=1$ 
           & 2.1035& 2.1035& 8.6379 &  4.0674  &  4.0674  & 0.8942   \\
& $2$ &  0.1902 &    0.1902& 2.4317 &  0.3369  &  0.3369  & 0.08141  \\
& $3$ & 0.02988  &  0.02988& 1.0792 &  0.04755 &  0.04755 & 0.01285  \\
& $4$ & 0.005858 & 0.005858& 0.5685 &  0.008311&  0.008311& 0.002523 \\ 
& $5$ &  0.001295 &0.001295& 0.3272 &  0.001630&  0.001630& 0.0005570  \\[1mm]
\hline 
\rule[0mm]{0mm}{2.5ex} ${\cal B}$ & $n=1$ 
     & $44.999\pm 0.002$   &  $23.342\pm 0.002$   & $172.778\pm 0.007$& $-9.888 \pm 0.006$  &$63.976 \pm 0.006$  & $12.689 \pm 0.135$  \\
&$2$ & $6.2595 \pm 0.0004$ &  $3.0899\pm 0.0008$  & $81.184\pm 0.005$ & $4.5354\pm 0.0005$  &$14.719 \pm 0.001$  & $1.2929\pm 0.0001$  \\
&$3$ & $1.1284 \pm 0.0001$ &  $0.4576\pm 0.0002$  & $42.771\pm 0.003$ & $0.6672\pm 0.0001$  &$2.7646 \pm 0.0003$ & $0.19901\pm 0.00005$\\
&$4$ & $0.24637\pm 0.00003$&  $0.08363\pm 0.00003$& $25.816\pm 0.002$ & $0.10688\pm 0.00002$&$0.60690\pm 0.00006$& $0.03777\pm 0.00001$\\
&$5$ & $0.06009\pm 0.00001$&  $0.01759\pm 0.00001$& $16.873\pm 0.001$ & $0.01865\pm 0.00001$&$0.14713\pm 0.00002$& $0.00804\pm 0.00001$\\[1mm]
\hline
\rule[0mm]{0mm}{2.5ex} ${\cal C}$ & $n=1$
     & $867.60 \pm 21.19$&$204.55\pm 19.79$ & $3212.2\pm 88.7$& $46.2\pm 47.8$  & $655.5\pm 72.0$ & $110.6 \pm 9.3$\\
&$2$ & $172.80 \pm 1.22$&$ 38.93\pm  1.12$ & $2220.9\pm 12.0$& $40.7 \pm 1.9$    & $337.8\pm 2.5$  & $13.84 \pm 0.62$\\
&$3$ & $35.38 \pm  0.26$&$ 4.626\pm 0.196$ & $1296.6\pm 6.7$ & $5.61 \pm 0.26$   & $73.34\pm 0.36$ & $1.892 \pm 0.107$\\
&$4$ & $8.13 \pm   0.04$&$ 0.445\pm 0.051$ & $843.1 \pm 3.9$ & $0.591 \pm 0.051$ & $17.06\pm 0.07$ & $0.274 \pm 0.027$\\
&$5$ & $2.109\pm  0.009$&$0.0367\pm 0.0121$& $585.0 \pm 3.2$ & $-0.273\pm 0.011$ & $4.29\pm  0.02$ & $0.0585 \pm 0.005$\\[1mm]
\hline
\end{tabular}
\caption{Contributions to event shape moments at LO, NLO, NNLO.\label{tab:mom}}
\end{sidewaystable} 
\clearpage

\FIGURE[htb]{ 
\epsfig{file=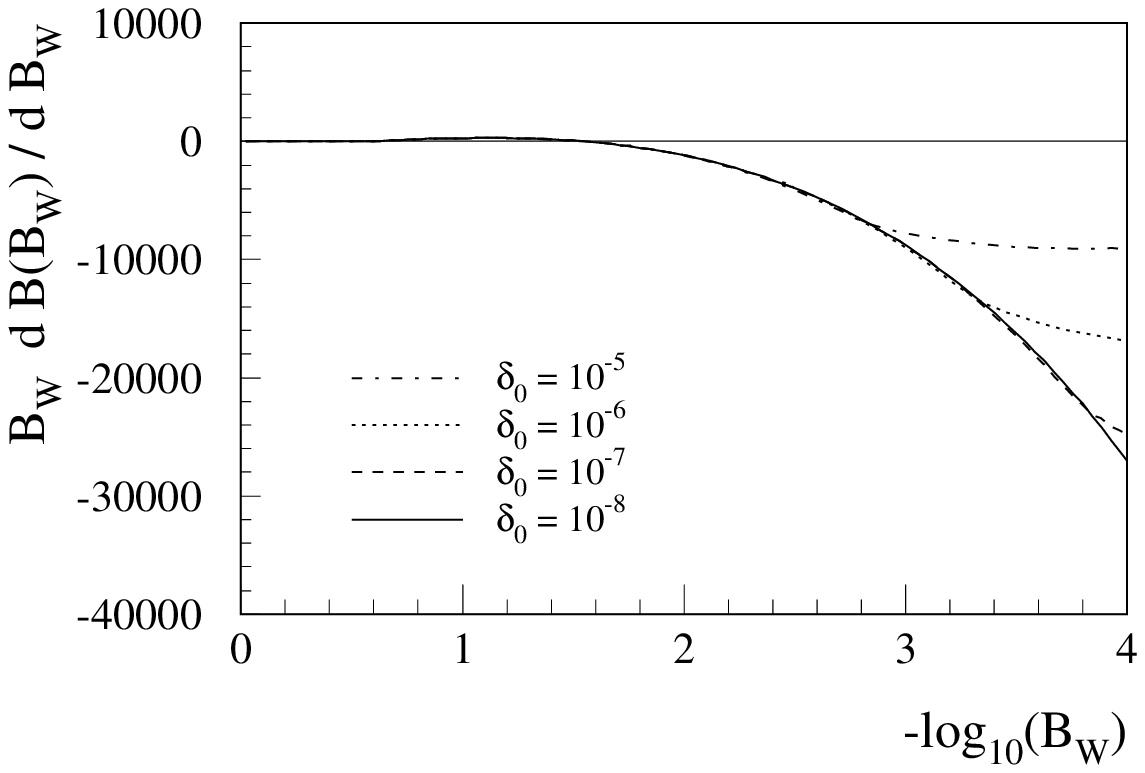,angle=-90,width=9cm}
\label{fig:bwnlo}
\caption{The NLO contribution to the wide jet broadening distribution 
$B_W\,dB(B_W)/dB_W$
evaluated for different technical cut-offs $\delta_0$ on  
all phase space variables.}
}

\FIGURE[htb]{ 
\epsfig{file=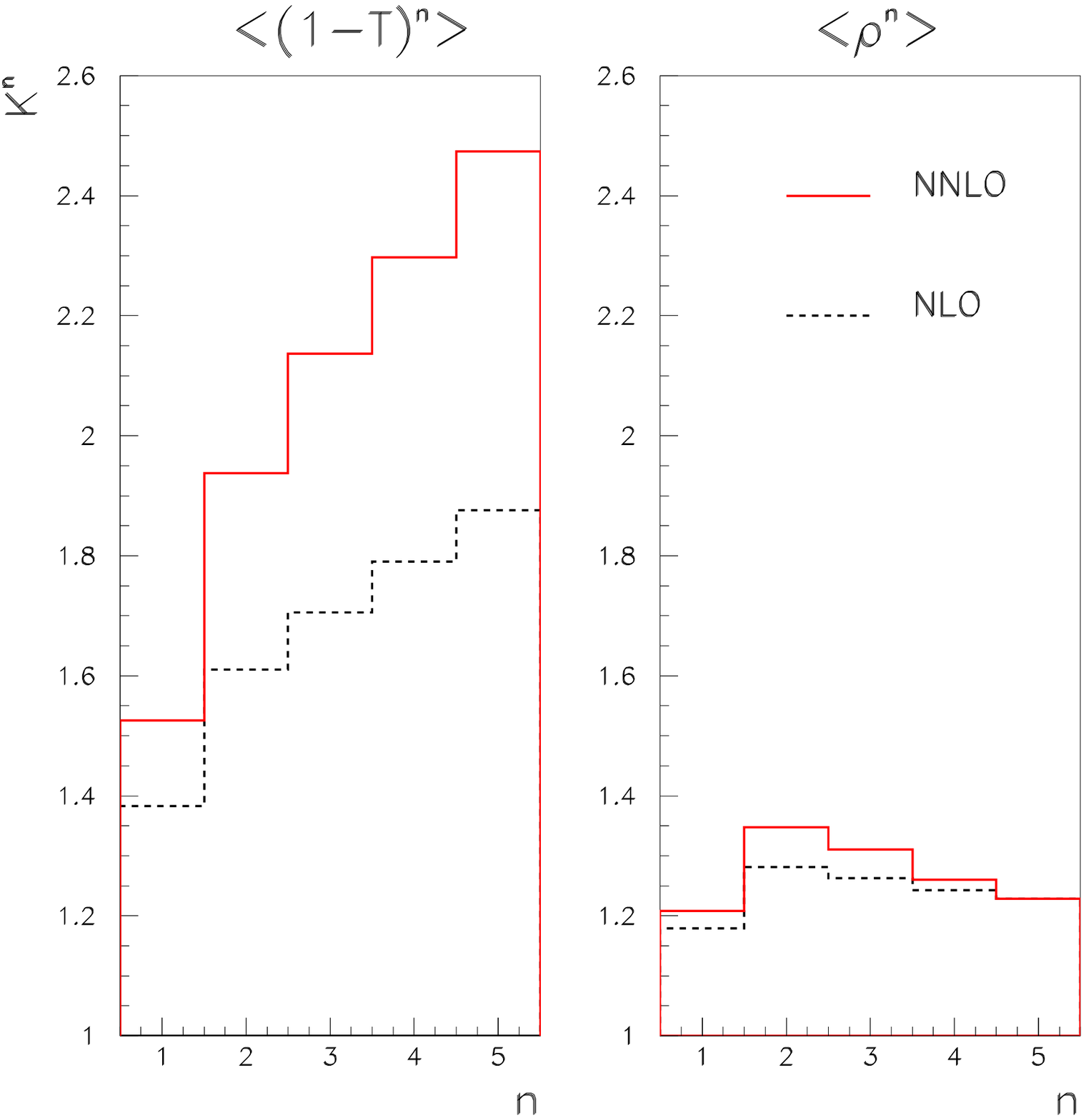,width=12cm}
\label{fig:TMH}
\caption{The ratios $K_{\rm{NNLO}}$ ($K_{\rm{NLO}}$)  
 of the NNLO (NLO) corrections relative to LO for the first five moments of the thrust and heavy mass distributions evaluated at  $\mu = Q$ with  
$\alpha_s=0.124$.}
}

\FIGURE[htb]{ 
\epsfig{file=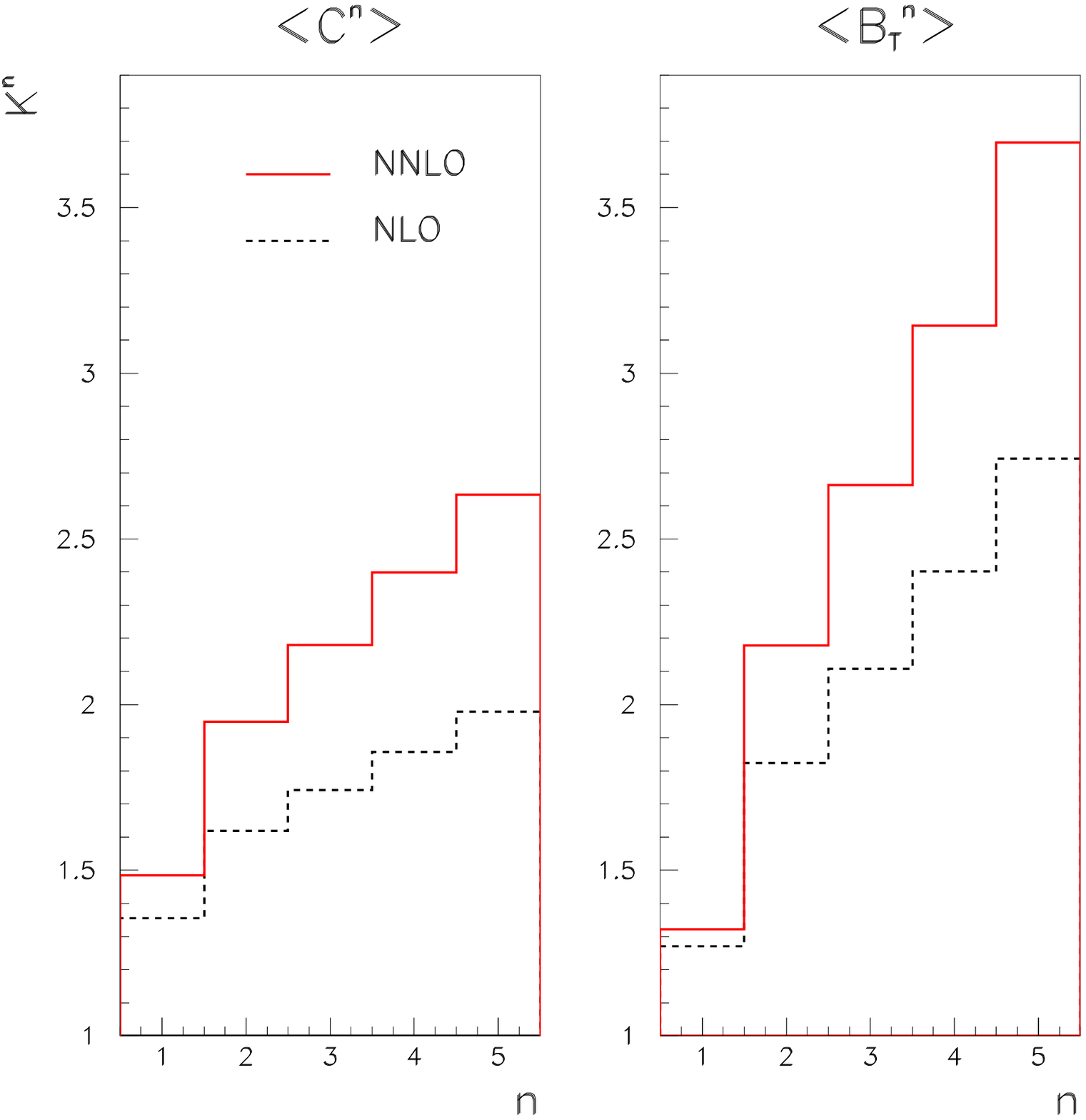,width=12cm}
\label{fig:CBT}
\caption{The ratios $K_{\rm{NNLO}}$ ($K_{\rm{NLO}}$)  
 of the NNLO (NLO) corrections relative to LO for the first five moments of the $C$-parameter and total jet broadening distributions evaluated at  $\mu = Q$ with $\alpha_s=0.124$.}
}

\FIGURE[htb]{ 
\epsfig{file=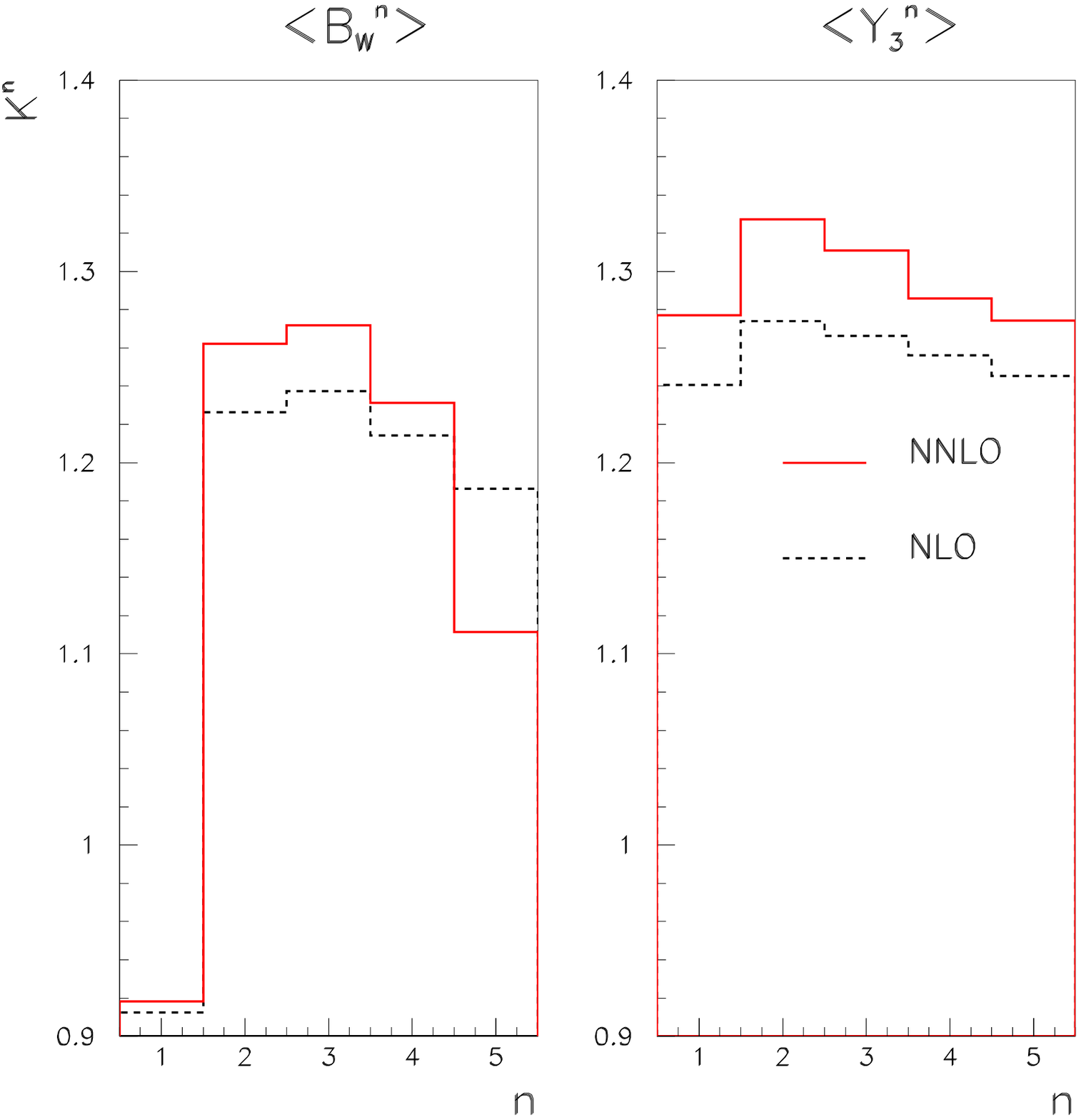,width=12cm}
\label{fig:BWY3}
\caption{The ratios $K_{\rm{NNLO}}$ ($K_{\rm{NLO}}$)  
 of the NNLO (NLO) corrections relative to LO for the first five moments of the wide jet broadening and $Y_3$ distributions evaluated at $\mu = Q$ with  
$\alpha_s=0.124$.}
}

\FIGURE[htb]{ 
\epsfig{file=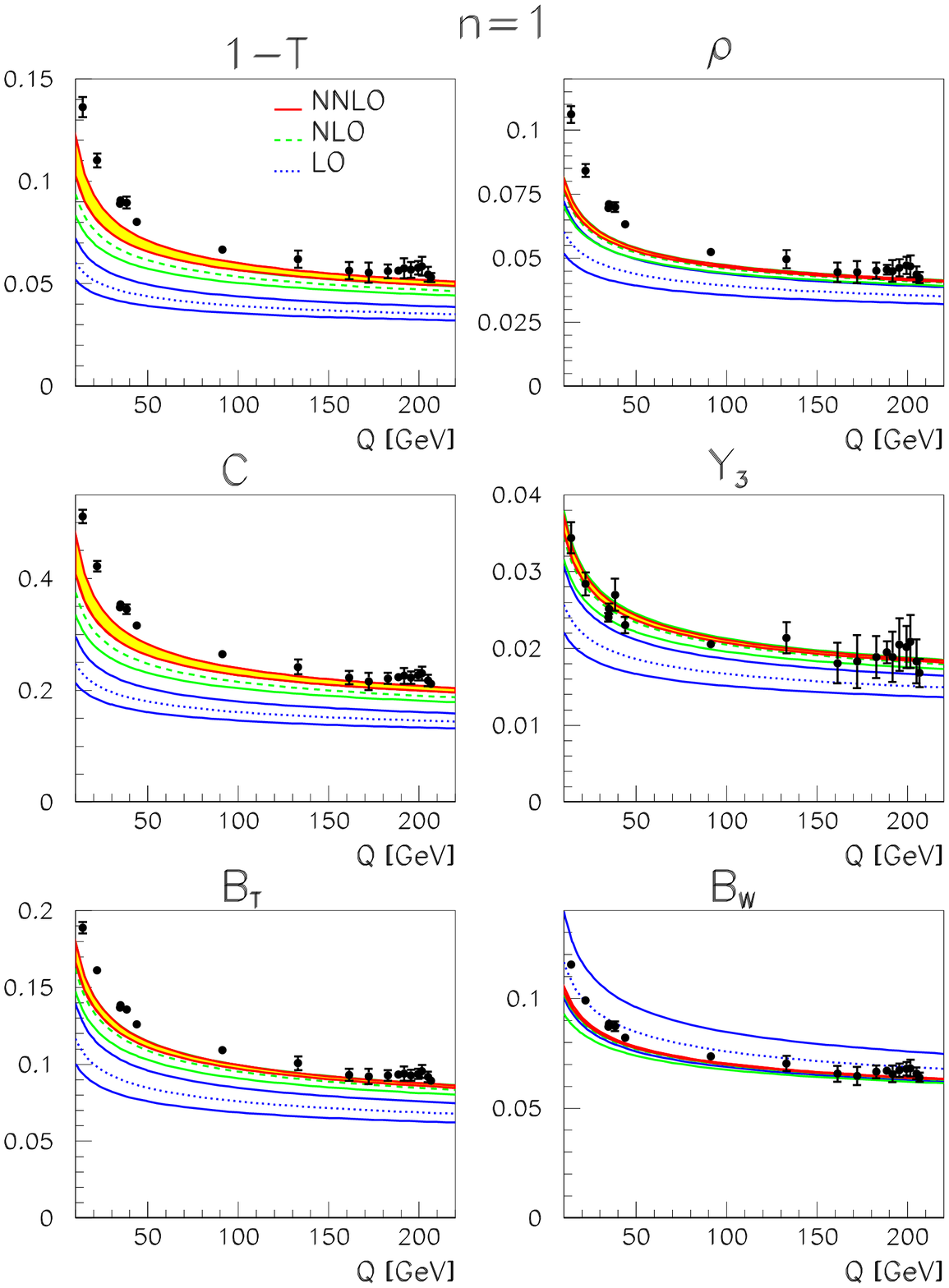,width=18cm}
\label{fig:tn1}
\caption{First moments of six event shape variables. 
The bands show the scale variations obtained by 
using $\mu=2Q$ and $\mu=Q/2$.
The data are 
from the JADE and OPAL experiments, taken from \cite{Pahl:2007zz}.}
}
\FIGURE[htb]{ 
\epsfig{file=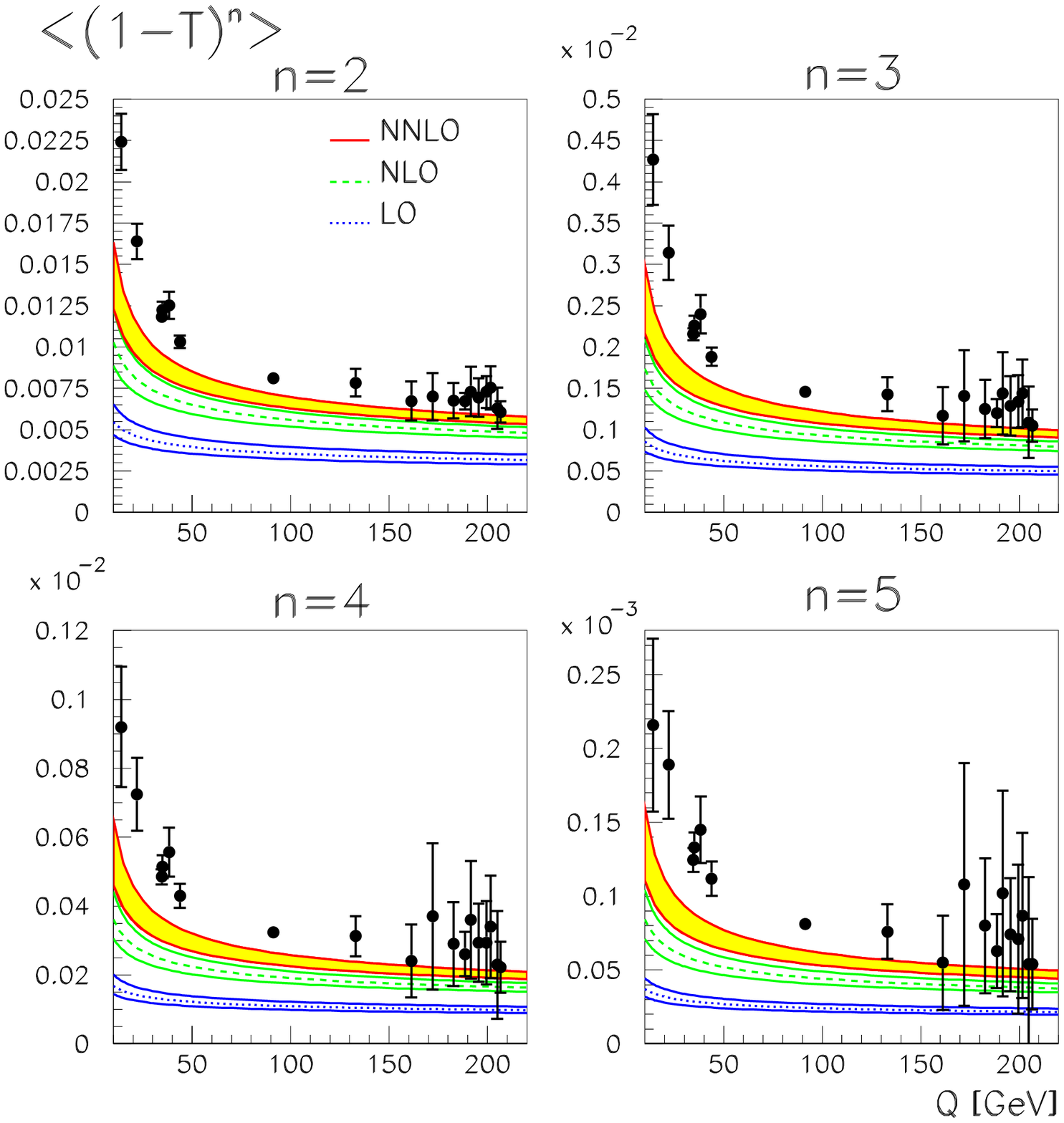,width=15cm}
\label{fig:tn2to5}
\caption{Perturbative corrections to the higher moments of $1-T$. 
The bands show the scale variations obtained by 
using $\mu=2Q$ and $\mu=Q/2$.
The data are taken from \cite{Pahl:2007zz}.}
}
\FIGURE[htb]{ 
\epsfig{file=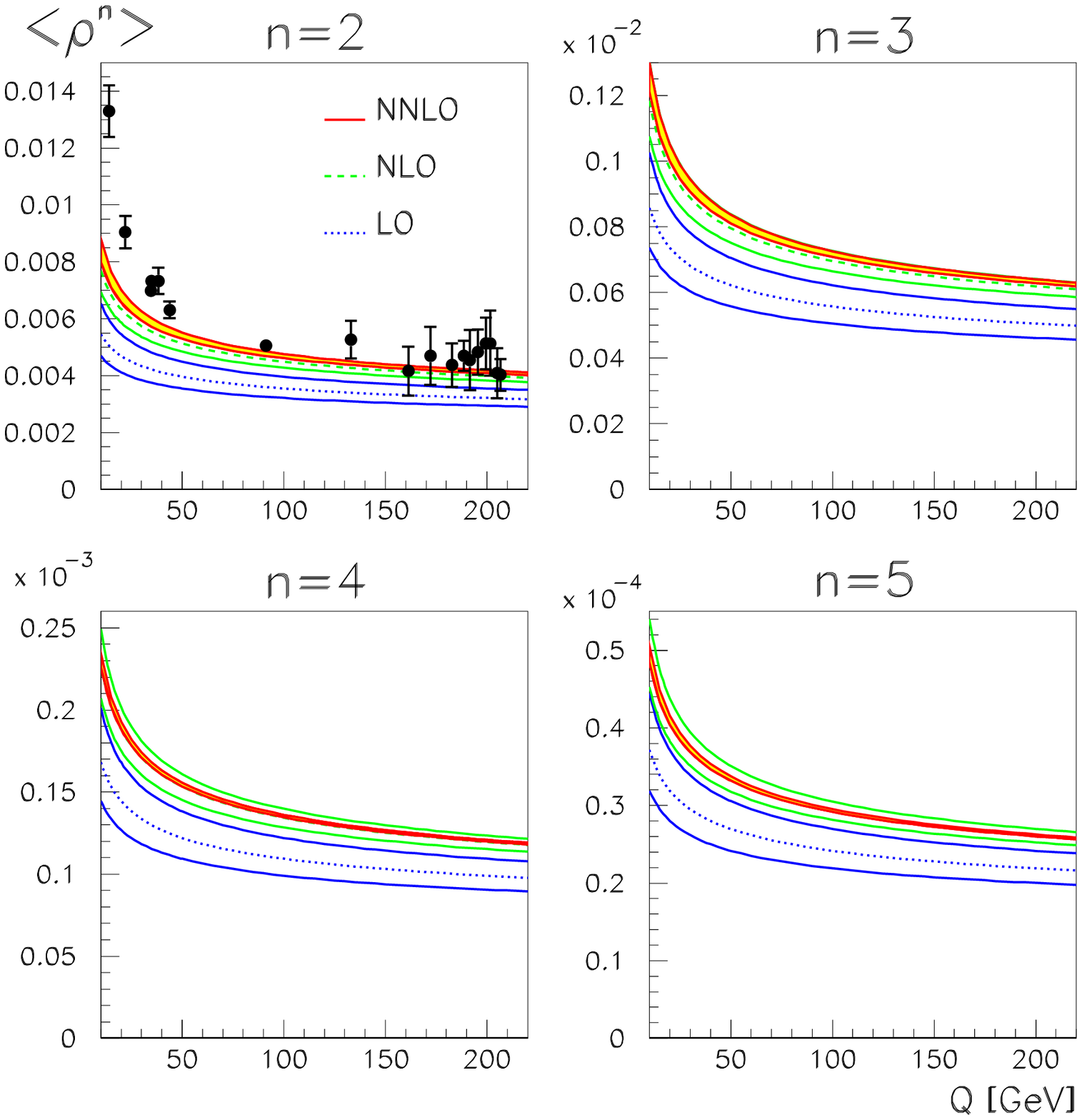,width=15cm}
\label{fig:rhon2to5}
\caption{Perturbative corrections to the higher moments of the 
normalised heavy jet mass $\rho$. 
The bands show the scale variations obtained by 
using $\mu=2Q$ and $\mu=Q/2$.
The data are taken from \cite{Pahl:2007zz}, which only 
contains data for the moments of $M_H=\sqrt{\rho}$, such that no data are 
shown for $n=3,4,5$.}
}

\FIGURE[htb]{ 
\epsfig{file=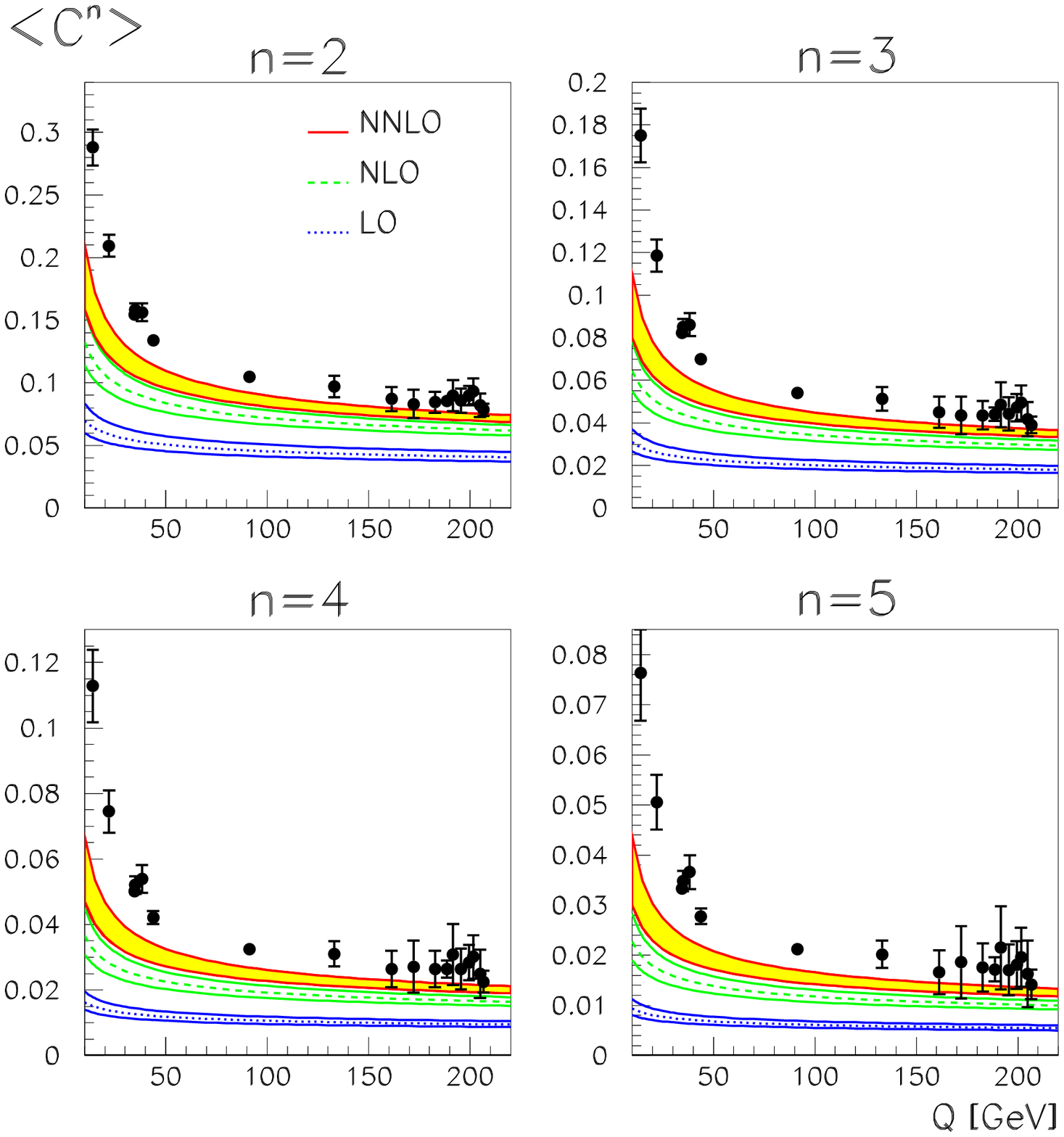,width=15cm}
\label{fig:cn2to5}
\caption{Perturbative corrections to the higher moments of the $C$-parameter. 
The bands show the scale variations obtained by 
using $\mu=2Q$ and $\mu=Q/2$.
The data are taken from \cite{Pahl:2007zz}.}
}
\FIGURE[htb]{ 
\epsfig{file=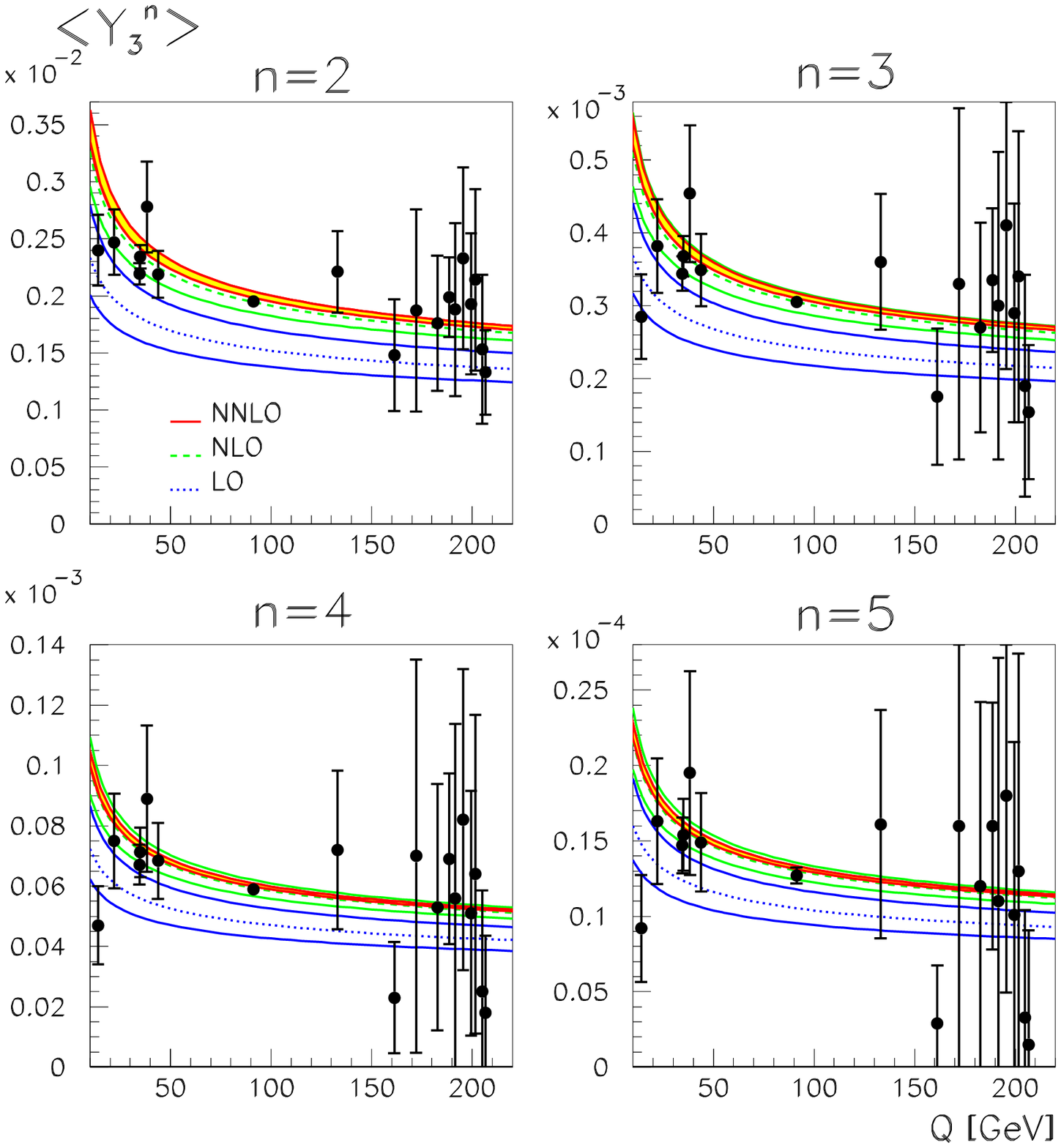,width=15cm}
\label{fig:y3n2to5}
\caption{Perturbative corrections to the higher moments of the 
2-jet to 3-jet transition parameter $Y_3$. 
The bands show the scale variations obtained by 
using $\mu=2Q$ and $\mu=Q/2$.
The data are taken from \cite{Pahl:2007zz}.}
}

\FIGURE[htb]{ 
\epsfig{file=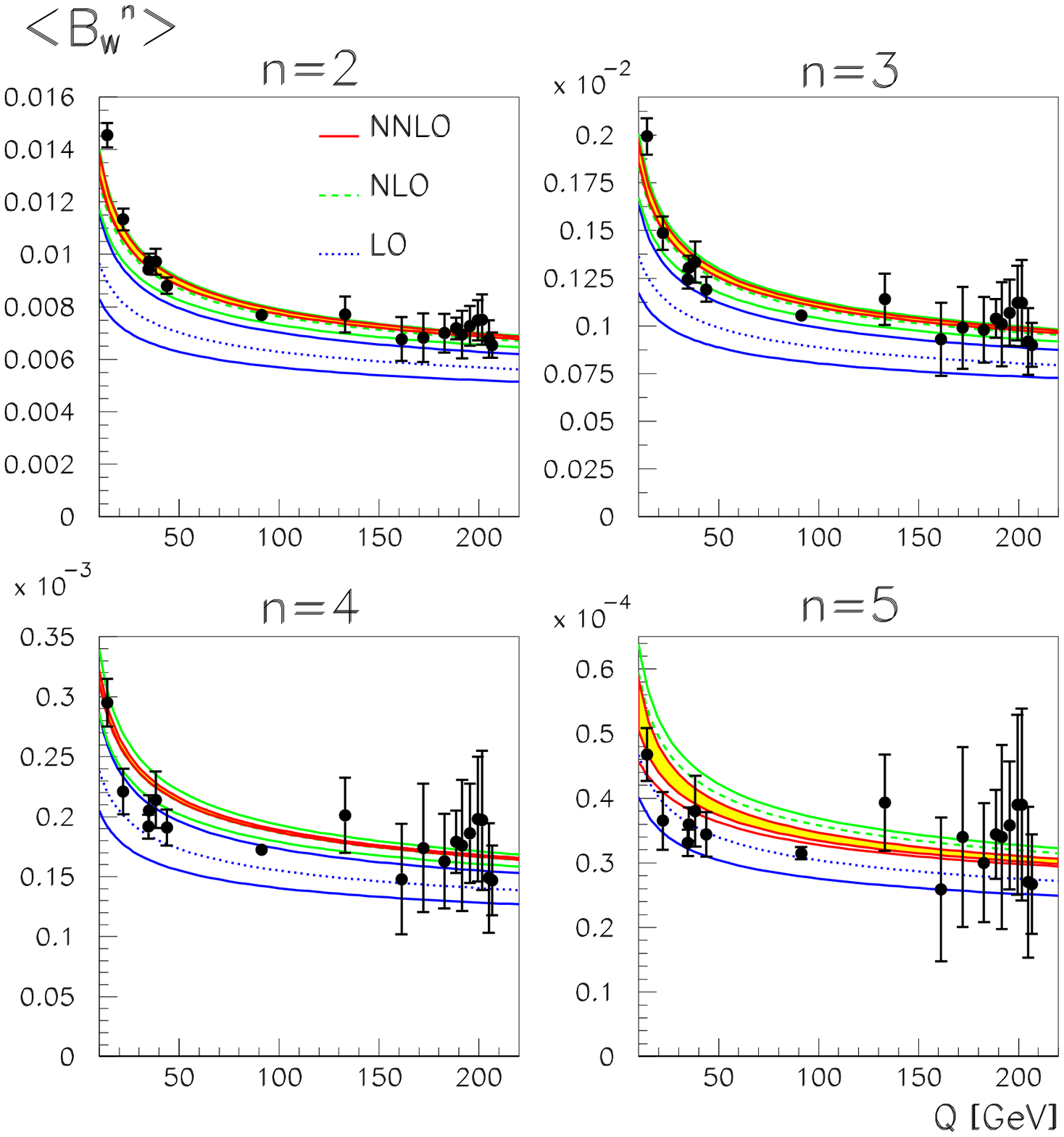,width=15cm}
\label{fig:bwn2to5}
\caption{Perturbative corrections to the higher moments of the 
wide jet broadening $B_W$. 
The bands show the scale variations obtained by 
using $\mu=2Q$ and $\mu=Q/2$.
The data are taken from \cite{Pahl:2007zz}.}
}
\FIGURE[htb]{ 
\epsfig{file=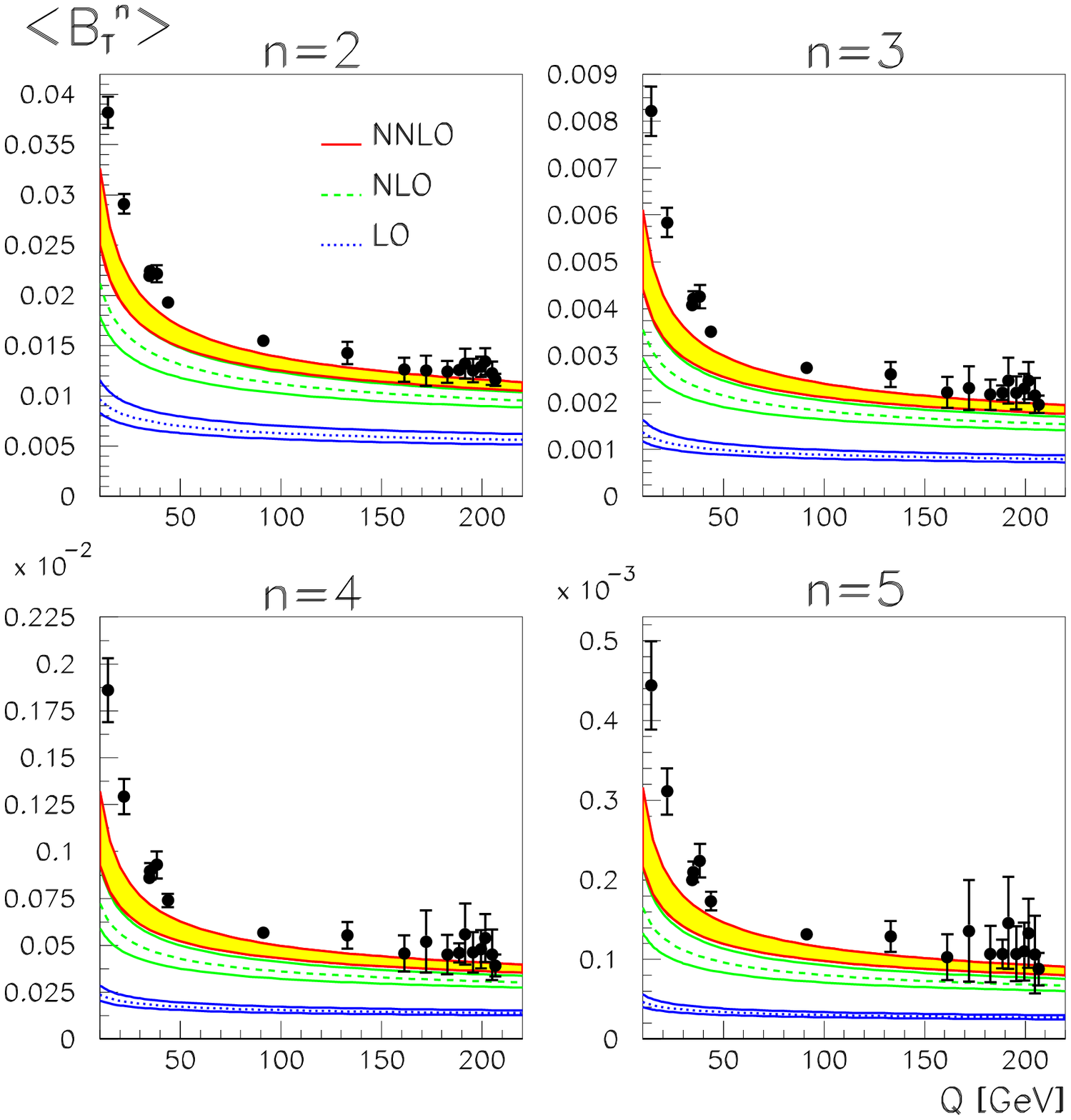,width=15cm}
\label{fig:btn2to5}
\caption{Perturbative corrections to the higher moments of the 
total jet broadening $B_T$. 
The bands show the scale variations obtained by 
using $\mu=2Q$ and $\mu=Q/2$.
The data are taken from \cite{Pahl:2007zz}.}
}

\end{document}